\documentclass{article}
\usepackage[T1]{fontenc}
\usepackage[utf8]{inputenc}
\usepackage{ismir} 
\usepackage{amsmath,amssymb,cite,url}
\usepackage{graphicx}
\usepackage{color}
\usepackage{bbm}
\usepackage{xcolor}
\usepackage{booktabs}
\usepackage{multirow}
\usepackage{lilyglyphs}

\title{PeakNetFP: Peak-based Neural Audio Fingerprinting\\Robust to Extreme Time Stretching}

  \multauthor
  {Guillem~Cort\`{e}s-Sebasti\`{a}$^{1 3}$\hspace{1cm} Benjamin~Martin$^2$ \hspace{1cm} Emilio~Molina$^1$}
  {{\bf Xavier~Serra$^3$ \hspace{1cm} Romain~Hennequin$^2$}\\
  $^1$ BMAT Licensing S.L., Barcelona, Spain\\
  $^2$ Deezer Research, Paris, France \\
  $^3$ Music Technology Group, Universitat Pompeu Fabra, Barcelona, Spain\\
  {\tt\small research@bmat.com, research@deezer.com}
  }

\def\authorname{G. Cort\`{e}s-Sebasti\`{a}, B. Martin, E. Molina, X. Serra, R. Hennequin}

\begin{document}

\maketitle

\begin{abstract}
This work introduces \emph{PeakNetFP}, the first neural audio fingerprinting (AFP) system designed specifically around spectral peaks. This novel system is designed to leverage the sparse spectral coordinates typically computed by traditional peak-based AFP methods. \emph{PeakNetFP} performs hierarchical point feature extraction techniques similar to the computer vision model \emph{PointNet++}, and is trained using contrastive learning like in the state-of-the-art deep learning AFP, \emph{NeuralFP}. This combination allows \emph{PeakNetFP} to outperform conventional AFP systems and achieves comparable performance to \emph{NeuralFP} when handling challenging time-stretched audio data. In extensive evaluation, \emph{PeakNetFP} maintains a Top-1 hit rate of over 90\% for stretching factors ranging from 50\% to 200\%. Moreover, \emph{PeakNetFP} offers significant efficiency advantages: compared to \emph{NeuralFP}, it has 100 times fewer parameters and uses 11 times smaller input data. These features make \emph{PeakNetFP} a lightweight and efficient solution for AFP tasks where time stretching is involved. Overall, this system represents a promising direction for future AFP technologies, as it successfully merges the lightweight nature of peak-based AFP with the adaptability and pattern recognition capabilities of neural network-based approaches, paving the way for more scalable and efficient solutions in the field.
\end{abstract}

\section{Introduction}\label{sec:introduction}

Audio Fingerprinting (AFP) is the MIR task of identifying audio recordings within a database of reference tracks. Early AFP systems date back twenty years, with Shazam \cite{wang2003industrial} and Philips \cite{haitsma2002highly} systems. Since then AFP has been extensively studied for various use cases, such as query-by-example \cite{wang2003industrial}, integrity verification \cite{gomez2002mixed}, content-based copy detection \cite{ouali2014robust}, DJ-set monitoring \cite{sonnleitner2016landmark}, or high specific audio retrieval \cite{chang2021neural}. Peak-based AFP systems have a long trajectory in the field and multiple works use this approach to enhance their robustness to pitch shifting and time stretching \cite{six2014panako}, background music identification \cite{kim2024background}, or to create a lightweight AFP that can run in embedded systems \cite{six2020olaf}. These algorithms are based on extracting and linking salient spectral peaks computed from time-frequency representations. These are mature, production-ready systems that do not require training, and can scale to industrial levels in which databases consist of millions of references \cite{shazam2009patent,US10657174B2}. Thus, companies with massive data catalogs rely on them for content identification \cite{akesbi2022audio}.

Representation learning systems, such as \emph{Now Playing} \cite{gfeller2017now} or \emph{NeuralFP} \cite{chang2021neural}, recently emerge as novel approaches that leverage Contrastive Learning (CL) and Convolutional Neural Networks (CNNs) to learn the similarities between a distorted audio clip and its corresponding reference track. They are designed to perform highly sensitive audio retrieval, capable of matching short segments, and significantly outperform traditional peak-based methods under challenging conditions. This is due to their ability to capture more complex and nuanced features from the data, making them more robust to various types of distortions and noise that traditional methods struggle with \cite{chang2021neural,cortes2022BAF}.

This comes at the expense of requiring large computational resources, dense input data, model training, and GPU computing, which might not be suitable for some applications. In industrial solutions, these requirements may be hard to overcome, and peak-based features are still considered as a viable alternative \cite{wang2006shazam,akesbi2022audio,US10418051B2,US10657174B2}. Practically, it is common that audio features have to be computed on a client device, then uploaded to a server to perform identification against a reference database. In such conditions, requiring dense spectrograms as audio features significantly increases the amount of data to be uploaded compared to sending only sparse spectral peaks. When fingerprint generation is considered to run client-side, it may be more complex and battery intensive to run inference on trained models as compared to simpler rule-based peak extraction algorithms, especially for query-by-example applications \cite{wang2006shazam} where client devices are generally variable in specifications (e.g. smartphones). An alternative setup is fully in-device audio identification \cite{gfeller2017now}, although this generally implies even more restrictive computational requirements on the device in terms of memory footprint and limits the database size. Additionally, music copyrights owners are often reluctant to share any dense representation that could be either inverted or used for other tasks than fingerprinting, and are more inclined to compute sparse targeted features that carry less information and can hardly be used for anything else than what they were designed to (with rare exceptions \cite{pfister2024listening}). Finally, as peak-based AFP has been used extensively by industrial systems, it is relevant for private companies to leverage such large datasets of pre-computed spectral peaks for neural audio fingerprinting approaches \cite{wang2003industrial,akesbi2022audio}.
For these reasons, in this work we propose to keep the traditional peak-based features as input, and use them in a modern neural approach.

In this first publication on a neural sparse peak-based model, we choose to focus our study on time stretching in extreme conditions, which has been underexplored in the literature. Time stretching is an audio processing technique that alters the tempo of a track without changing its pitch. This method is commonly used by DJs to synchronize the tempo of different songs within a mix or to create remixes that are either slowed down or sped up \cite{schwarz2018unmixdb}. In challenging situations such as mash-ups, blends, or licensing circumvention attempts, time stretching happens in complex identification situations where severe tempo modifications are used on short excerpts, making them very hard to be automatically identified \cite{sonnleitner2017thesis}.

The main contribution of this work is to introduce a novel AFP system operating with lightweight spectral peaks as input, but grounded in a representation learning approach and evaluated in the context of time stretching. Specifically, our model \emph{PeakNetFP} applies contrastive learning to learn fingerprints from sparse spectral peaks input, leveraging the hierarchical point set learning algorithm \emph{Pointnet++} \cite{qi2017pointnetplusplus}. It is designed to exhibit the good performance of neural state-of-the-art approaches while keeping memory footprint low thanks to sparse input. To our knowledge, this is the first attempt at combining traditional peaks and representation learning for audio fingerprinting, and the first time a point-cloud network is used for AFP.
As a subsequent contribution, we evaluate \emph{PeakNetFP} alongside the SOTA algorithm on time stretching, \emph{QuadFP} \cite{sonnleitner2016robust}, which is a peak-based approach, and \emph{NeuralFP} \cite{chang2021neural}, the SOTA neural audio fingerprinting, in a new scenario to it. We finally show that \emph{PeakNetFP} 
 achieves performance close to the SOTA method \emph{NeuralFP}, despite using 100 times fewer parameters and 11 times smaller input data than the latter.

In section \ref{sec:relatedwork} we summarize the works relevant to this publication. In section \ref{sec:methodology}, we describe the hierarchical peak set feature extraction as well as the contrastive representation learning framework at the core of \emph{PeakNetFP}. Finally, in section \ref{sec:results} we present its evaluation in the context of extreme time stretching and show how it compares to the peak-based time stretching baseline \emph{QuadFP}, and to the spectrogram-based SOTA model \emph{NeuralFP}.
\emph{PeakNetFP} code, dataset, and model are open and available\footnote{\url{https://github.com/guillemcortes/peaknetfp}}.

\section{Related Work}
\label{sec:relatedwork}

Over the past two decades, the research community has worked to advance audio fingerprinting systems for multiple use cases. Some of these innovations include wavelets \cite{baluja2008waveprint} for noise resilience, constant Q-transform \cite{fenet2011scalable} or Fundamental Frequency Map \cite{son2020robust} for pitch-shifting robustness, and cosine filters \cite{ramona2013audioprint} for broadcast monitoring, to name a few.

We can classify AFP methods into three broad categories: local descriptors-based \cite{son2020robust,Haitsma2001,baluja2008waveprint,ramona2013audioprint,ouali2014robust,anguera2012mask,dupraz2010robust,agarwaal2023robust}, peak-based \cite{wang2003industrial,six2014panako,sonnleitner2016robust,sonnleitner2017thesis,dupraz2010robust,fenet2011scalable,sonnleitner2014quad,malekesmaeili2014local,lee2015audio}, and neural audio fingerprints \cite{gfeller2017now,yu2020contrastive,chang2021neural,singh2022attention,wu2022asymetric,singh2023simultaneously,fujita2024afphrr}. Peak-based fingerprints started with Shazam's algorithm \cite{wang2003industrial}, which set the basis of spectral peak pairs linking to form hashes that are robust to noise. Then, Six \& Leman proposed linking triplets to obtain robustness to time and frequency modifications in \emph{Panako} \cite{six2014panako}, although it is not suitable for short queries or extreme time stretching since it was designed for content deduplication of audio collections of old recordings that were digitalized by replaying. Similarly, Sonnleitner \& Wilder proposed \emph{QuadFP} \cite{sonnleitner2016robust}, which adapts blind astrometry research \cite{lang2010astrometry} to build quadruplets of peaks and generate hashes robust to significant time and frequency modifications \cite{sonnleitner2017thesis}. Other peak-based AFP works \cite{dupraz2010robust,fenet2011scalable,sonnleitner2014quad,malekesmaeili2014local,lee2015audio} have also studied how to improve the robustness to time and frequency modifications. Peak-based traditional methods perform well even in the presence of alterations such as noise, compression, or reverberation, for instance. They generally produce lightweight hashes that can be efficiently indexed into lookup tables, which makes them scalable to hundreds of thousands or even tens of millions of music pieces. Additionally, such methods do not require training or accelerated computing hardware.

\subsection{QuadFP}\label{subsec:quadfp}
However, such traditional methods significantly underperform in the presence of extremely challenging scenarios, like in the case of strong time stretching \cite{sonnleitner2016robust}. \emph{QuadFP} stands out as one of the most advanced peak-based AFP in that regard. Designed to be robust to time and frequency modifications, its core innovation is the use of quadruple peak descriptors, which capture not only the position of each peak but also its relationship with neighboring peaks. Each quadruple describes a constellation of four peaks (local maxima) in the time-frequency domain, effectively encoding local patterns and relationships between peaks. This approach is more robust to noise and variations in audio content compared to other fingerprinting methods \cite{sonnleitner2016robust} such as \textit{Panako} \cite{six2014panako}.
Once the quadruple features are extracted, \emph{QuadFP} uses a hashing mechanism to map these descriptors to a database. In this publication, we use \emph{QuadFP} as the most advanced peak-based AFP baseline for robustness against time stretching. It also aligns with the use case of this study, which restricts the input data to spectral peaks. Our objective is to show how much a neural network can improve the best traditional system for time stretching.

\subsection{NeuralFP}\label{subsec:neuralfp}
In the last decade, neural networks have been successfully used for AFP. In 2017, Google presented the first neural AFP system \emph{Now Playing} \cite{gfeller2017now}. It was designed to run on mobile devices with datasets of limited size while featuring high robustness to noise. A very recent approach, GraFPrint \cite{bhattacharjee2025grafp}, leverages the structural learning capabilities of Graph Neural Networks (GNNs) to generate robust fingerprints from time-frequency representations. As opposed to our method, rather than using sparse spectral peaks, GraFPrint extracts from spectrograms localization-aware, low-dimensional features using a convolutional encoder. Other neural AFP systems \cite{yu2020contrastive,chang2021neural,singh2022attention,wu2022asymetric,singh2023simultaneously,fujita2024afphrr} use targeted augmentations in a contrastive learning framework to achieve robustness to noise, reverberation, echo, and other distortions.
Among them, \emph{NeuralFP} \cite{chang2021neural} stands out as being the only fully replicable neural AFP. The implementation is open-source, with a public dataset and model weights. \emph{NeuralFP} is also lighter than other proposed models such as transformer-based AFPs \cite{singh2022attention,singh2023simultaneously}.
 
 For these reasons, we use \emph{NeuralFP} as a foundation for the development of \emph{PeakNetFP}. \emph{NeuralFP} leverages contrastive learning to achieve high-sensitive audio retrieval, employing a convolutional encoder to extract meaningful features from mel-spectrograms. In this paper, we propose to reuse \emph{NeuralFP}'s contrastive learning framework while changing input data from dense spectrograms to sparse peak-based features. The original \emph{NeuralFP} is then used as a reference model, showing what could be achieved when considering full spectrograms as input.

\subsection{AFP for time stretching}
Some fingerprinting methods have been designed to effectively handle time stretching, the focus of this work.

\emph{QuadFP} \cite{sonnleitner2016robust} appears as a milestone on this topic. Using a quadruple-based spectral peak grouping (see section \ref{subsec:quadfp}) combined with an asymmetric query-reference fingerprints configuration that maximizes the number of quads generated in queries, they report high precision and accuracy measures for multiple tempo modifications for 20 seconds queries. Their reference database consists of 100,000 tracks from Jamendo and they test 300 query tracks time-stretched with 13 stretching factors between 70\% and 130\%. In this experiment, they report an average accuracy of 92.9\% for 10-second queries, but 28.1\% average accuracy for 2.5-second queries. This shows how the performance collapses as the query length shrinks. We can expect this performance to be lower if fewer quads per second are used, a more likely scenario in an industrial environment. 
 
Yao et al. \cite{yao2018enhancing} use the same dataset as \cite{sonnleitner2016robust}. Experiments were done for 13 stretching factors between 70\% and 130\% and a query length of 20 seconds. They report similar performance to \emph{QuadFP} but with a 20\% drop in recall. SAMAF \cite{suarez2020samaf} reports that for different query lengths ranging from 1 to 6 seconds, they achieve over 80\% accuracy for mild stretching (0.9 and 1.1) but this collapses with severe stretching (0.5, 1.5), with less than 13\% of accuracy. Panako \cite{six2014panako} reports results for queries of 20, 40, and 60 seconds on a database of 30,000 songs. Less than a third of the queries are resolved correctly after a time stretching modification of 8\%, though. Son et al. \cite{son2020robust} achieve perfect precision for tempo modification in the range of 70\% to 130\%. However, their dataset only comprises 100 audio files and they query using the full audio length.
George \& Jhunjhunwala \cite{george2015scalable} propose to encode the features using only frequency information and thus making it independent of time, as opposed to \cite{wang2003industrial}, which encodes with respect to time. They test with tempo modifications in the $\pm$50\% range. They achieve over 97\% of accuracy but on a small dataset of 300 samples of 20 seconds each. Their algorithm is also not suitable for short queries.

\section{PeakNetFP}
\label{sec:methodology}

\begin{figure}[t]
    \centering
    \includegraphics[width=\columnwidth]{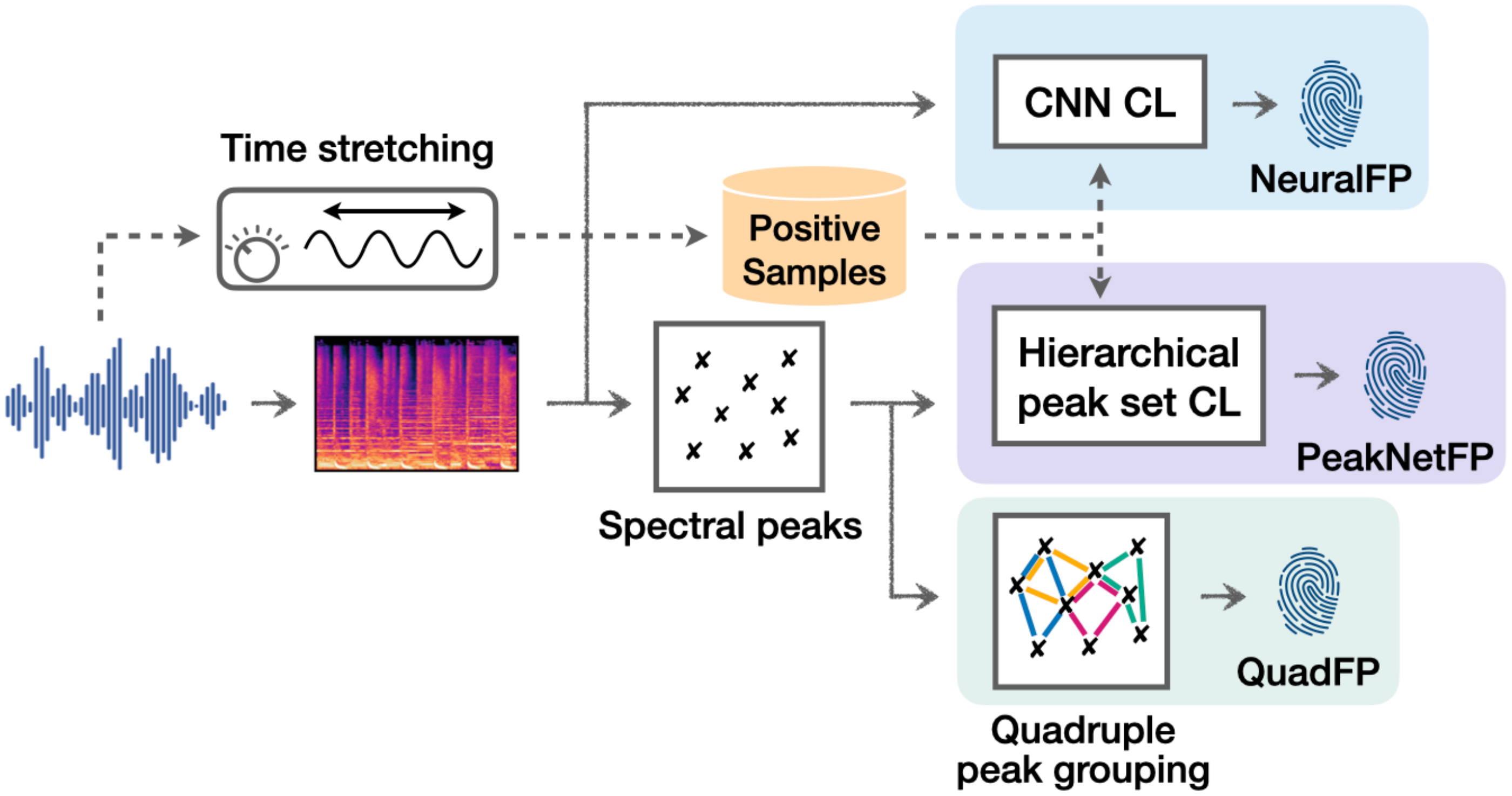}
    \caption{Considered AFP systems overview. From top to bottom: \emph{NeuralFP}, \emph{PeakNetFP} (ours), \emph{QuadFP}. Dashed lines represent extra data used for training. Our model \emph{PeakNetFP} learns features from the same input as \emph{QuadFP}, in the same contrastive learning framework as \emph{NeuralFP}.}
    \label{fig:afp-models}
\end{figure}

In this section, we describe our proposed model, \emph{PeakNetFP}, starting from the sparse input data through to the contrastive learning framework, highlighting the hierarchical peak set feature extraction process. Additionally, we introduce the dataset used for evaluation. Figure \ref{fig:afp-models} provides an overview of all the AFP systems considered, including our \emph{PeakNetFP}, the baseline \emph{QuadFP} \cite{sonnleitner2016robust}, and the SOTA AFP model \emph{NeuralFP} \cite{chang2021neural}.

\subsection{Sparse input data}
\label{subsec:sparse-input-data}

As we described in the introduction, our model is designed to handle sparse data in the form of 3-dimensional spectral peaks, as features extracted from a third-party traditional AFP system. Typically, such peaks represent a subset of the local maxima from the spectrogram, chosen based on criteria that identify the most salient ones \cite{wang2003industrial,sonnleitner2014quad,six2014panako}. In our system, we extract local maxima in the melspectrogram using 3x3 kernels and stride 1 as a proxy for traditional peak-based fingerprints for simplicity and to avoid biasing the results on other system criteria. This allows the neural network to learn which peaks are most relevant for matching or classification even though a more refined peak selection could help reduce the dimensionality of the input, improving computational efficiency. In practice, we select the 256 highest amplitude local maxima per each 1-second segment, ensuring that we capture the most prominent features within each window.

When working with peaks rather than continuous data points, it becomes challenging to inject locality into convolutional kernels, which typically rely on dense, grid-like input structures. Peaks create a sparse representation of data, much like how point clouds are used in computer vision tasks. This sparsity complicates the direct application of traditional convolutional methods, as there is no inherent neighborhood structure. However, approaches like \emph{PointNet} \cite{qi2016pointnet} and derived methods from the point cloud literature offer a potential solution by grouping local peaks and processing them similarly to how convolutions operate on spectrograms. By leveraging local relationships among peaks, we can capture meaningful patterns without requiring dense, continuous input.

\begin{figure}[t]
    \centering
    \includegraphics[width=\columnwidth]{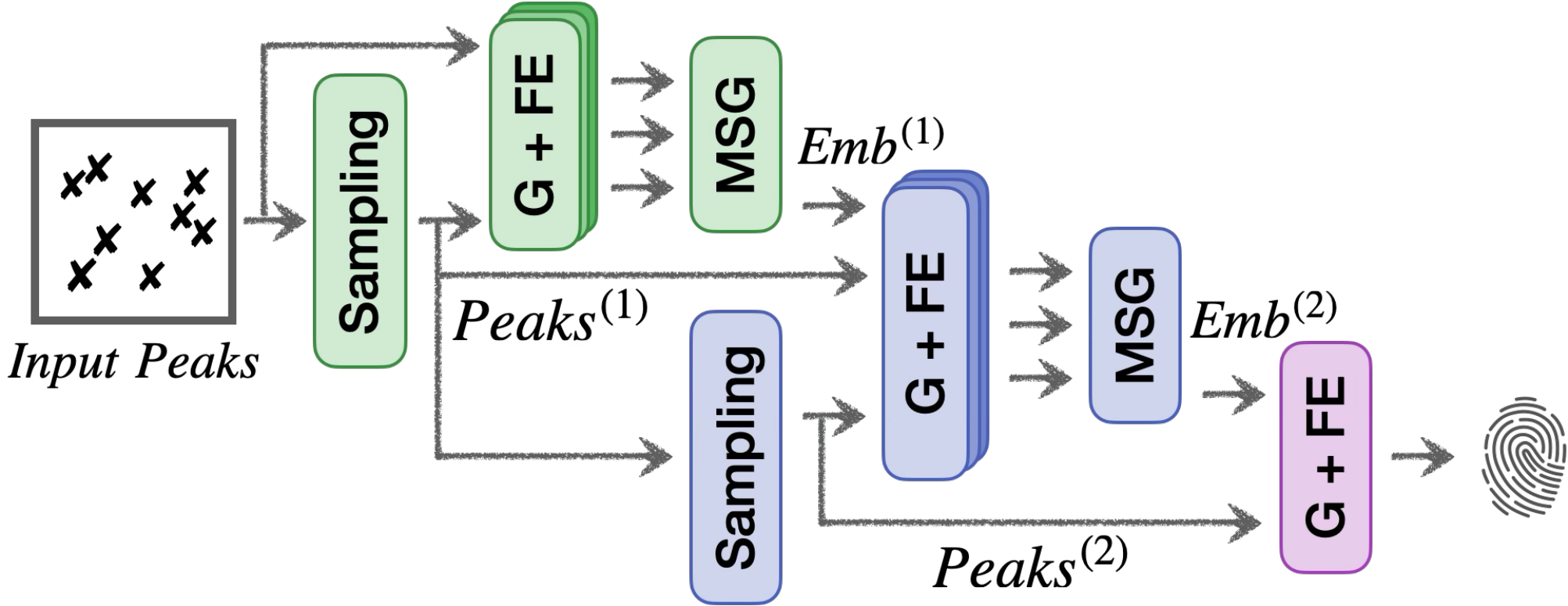}
    \caption{\emph{PeakNetFP} overview consisting of Sampling, Grouping + Feature Extraction (G+FE) and Multi-Scale feature Grouping layers (MSG).}
    \label{fig:peaknetfp}
\end{figure}

\subsection{Hierarchical peak set feature extraction}
\label{subsec:hierarchical-extraction}
\begin{figure}[t]
    \centering
    \includegraphics[width=\columnwidth]{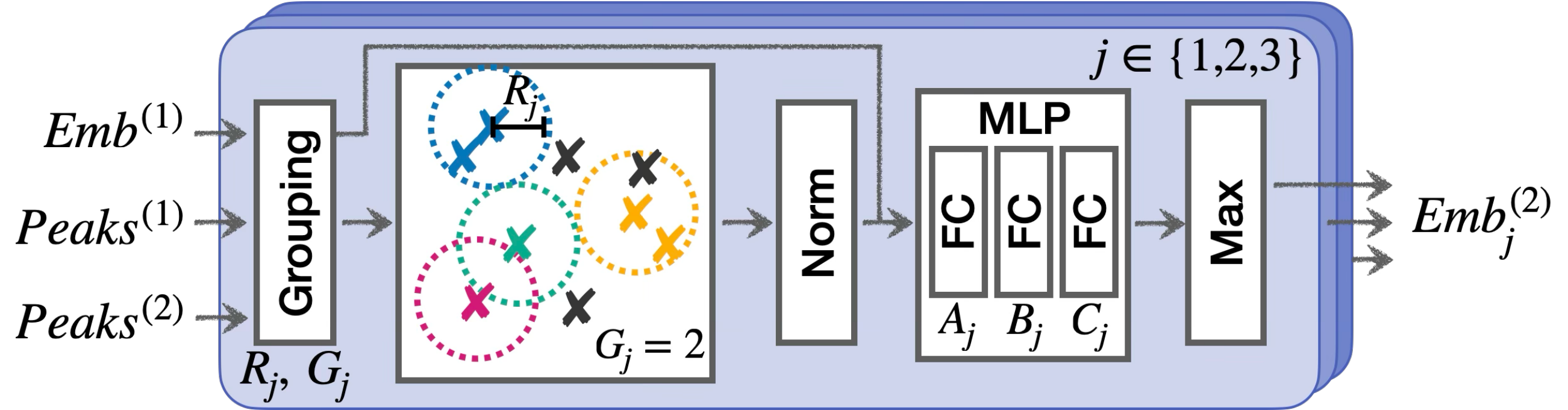}
    \caption{The Grouping + Feature Extraction block of the second layer ($i=2$). For each specific sublayer $j$, $R_j$ and $G_j$ are the radius and group size of the queryball, and $(A_j, B_j, C_j)$ are the dimensions of the MLP layers.}
    \label{fig:sa_block}
\end{figure}

Hierarchical \emph{PointNet}, or \emph{PointNet++} \cite{qi2017pointnetplusplus}, introduces a multi-level approach to capture both local and global features in sparse data, akin to the nested convolutions found in traditional CNNs. \emph{PointNet++} organizes points (peaks in our context) into hierarchical groupings, where local neighborhoods are progressively sampled and processed, similar to the way convolutions scan across dense data. This hierarchical structure enables \emph{PointNet++} to effectively learn both fine-grained and high-level features from sparse data. In \emph{PeakNetFP}, we incorporate the hierarchical peak set feature extraction from \emph{PointNet++} into the contrastive learning framework from \emph{NeuralFP}. Figure \ref{fig:peaknetfp} illustrates the architecture of \emph{PeakNetFP} and Figure \ref{fig:sa_block} the Grouping and Feature Extraction (G+FE) block of the second layer, represented in blue in Figure \ref{fig:peaknetfp}. 
The sparse peaks encoding starts with two Set Abstraction (SA) layers, represented in green and blue on Figure \ref{fig:peaknetfp}, which are responsible for grouping neighbouring peaks in a hierarchical way through Multi-Scale feature Grouping (MSG). Each layer $i$ operates in three key steps:

\vspace{0.2cm}
\noindent \textbf{(I) Sampling}: The $N^{(i)}$ peaks with greatest amplitudes are selected as anchor peaks that will be the center of the peak groups. 
This step helps control computational complexity as the network goes deeper.

\noindent \textbf{(II) Grouping + Feature Extraction block (G+FE)}: each block is made of 3 parallel layers, each layer $j$ comprising:

\textbf{(i) Grouping}: for each anchor peak, we select the $G^{(i)}_j$~closest~peaks withing radius $R^{(i)}_j$ using query balls to form local neighborhoods. These neighborhoods act as local receptive fields, similar to convolutional patches in CNNs.
The query ball is a crucial element because it allows precise control over the distance and radius for hierarchical search in the point cloud. Unlike traditional convolutions that rely on fixed grid structures, the query ball adapts to the irregular distribution of peaks by grouping them based on actual spatial proximity.

\textbf{(ii) Feature Extraction}: Within each neighborhood, an MLP with 3 layers of respective dimensions $A^{(i)}_j, B^{(i)}_j, C^{(i)}_j$ is applied to learn local features. The MLP aggregates features for each point and uses max-pooling to summarize them into a single vector of dimension $N^{(i)}\times C^{(i)}_j$ representing the local region.

\vspace{0.2cm}
\noindent \textbf{(III) Multi-scale feature grouping}: Features of all parallel extraction layers are concatenated to a single embedding of dimension $N^{(i)}\times(C^{(i)}_1+C^{(i)}_2+C^{(i)}_3)$.
This step allows the model to capture features at multiple scales simultaneously by using different neighborhood radii during the grouping stage, which helps consider both fine and coarse features. 

After each SA layer, the output is a smaller set of peaks with higher-dimensional feature vectors. These vectors are passed to the next SA layer, where the process repeats with a new sampling ($Peaks^{(2)}$ in Figure \ref{fig:peaknetfp}), further abstracting the data. As we move deeper into the network, the receptive fields become larger, allowing the network to capture broader contextual information while maintaining local details.
The last SA layer, represented in purple in Figure \ref{fig:peaknetfp}, is similar to a Grouping + Feature Extraction block, but where \textit{all} points are grouped together, forming a single 128-dimensional feature vector. This final vector thus encodes both local and global information about the peaks, and can be used as a fingerprint.

\subsection{Contrastive learning framework}
\label{subsec:contrastive-framework}
\emph{PeakNetFP} relies on the \emph{NeuralFP} contrastive learning framework \cite{chang2021neural} to learn fingerprints, which we describe in the following. It operates on 1-second windows with a 50\% overlap. It creates data pairs by applying time stretching to short audio snippets.
Each mini-batch $MB$ is formed by $N$ samples and $N$ augmented replicas of the same samples to generate positive pairs $x_i$ and $x_j$ so that $MB=\{x_i,x_j,...,x_{Ni}, x_{Nj}\}$ and $|MB|=2N$. 
NT-Xent loss \cite{chen2020simple} is chosen to maximize an agreement between positive pairs in a mini-batch $MB$. No explicit negative sampling is performed, thus, given a positive pair, the other $2(N-1)$ data points are to be treated as negative samples.
The NT-Xent loss for a given pair of embeddings \( \mathbf{z}_i \) and \( \mathbf{z}_j \) is defined as:
\begin{equation}
    \label{eq:loss-intra}
    l(i,j) = - \log \frac{\exp(a_{i,j} / \tau)}{\sum_{k=1}^{2N} \mathbbm{1}_{[k \neq i]} \exp(a_{i,k} / \tau)}
\end{equation}
Where \( a_{i,j} = \mathbf{z}_i^T \mathbf{z}_j \) for \(  i,j \in \{1,\dots,2N\} \). \( \tau \) is the temperature scaling factor for the softmax. Computing the Top-1 in the softmax function is equivalent to Maximum Inner Product Search (MIPS). \( \mathbbm{1}_{[k \neq i]} \) ensures that the summation excludes the anchor-positive pair. Then, the loss $\mathcal{L}$ averages $l$ across all positive pairs, both ($i,j$) and ($j,i$):
\begin{equation}
    \label{eq:loss-outer}
    \mathcal{L} = \frac{1}{2N} \sum_{k=1}^{N}l(2k-1,2k) + l(2k,2k-1)
\end{equation}
During retrieval, as in \cite{chang2021neural}, we retrieve 20 candidate segments from an Inverted File Product Quantization (IVFPQ) index built with \textit{Faiss} \cite{johnson2019billion}. Then, we perform a sequence matching in which each segment’s embedding is compared to candidate embeddings via inner product, and results are ordered based on this score.

\subsection{Dataset}
\label{subsec:dataset}
To develop \emph{PeakNetFP}, we use the same dataset as in \emph{NeuralFP}\cite{chang2021neural}
but change the augmentations to time stretching. The dataset consists in multiple audio files extracted from \texttt{fma\_medium} dataset \cite{defferrard2017fma} that comes with defined subsets, which we also use to train and test our models.
The \textit{Train} subset contains 10,000 30-second audio clips while \textit{Test-Query/DB} contains 500 30-second audio clips. To increase the reference set, we use \textit{Test-Dummy-DB}, which comprises 100,000 full tracks with an average length of 278 seconds each. This is useful for testing the scalability of the system. In the evaluation step, we use the same 2,000 segments selected randomly from the 500 clips as in the \emph{NeuralFP} evaluation.

During training, the stretching augmentations are performed at the spectrogram level, which we resize only on the time axis using bilinear interpolation. Unlike waveform-based stretching methods from \texttt{SOX}\footnote{\url{https://sourceforge.net/projects/sox/}}, this naive method integrates easily with the training pipeline and ensures that our model is not overfitting to any particularities of a specific model but learns to handle stretching. In testing, we use \texttt{SOX} to generate realistic queries from the DB tracks of \textit{Test-Query/DB} set. We generate queries for stretching factors 1.05, 1.1, 1.2, 1.4, 1.6, 1.8, and 2 that increase the tempo of the song, and their counterparts reducing the tempo 0.975, 0.95, 0.9, 0.8, 0.7, 0.6, and 0.5. Note that a stretching factor of 2 doubles the tempo while 0.5 halves it. This test set is publicly available in Zenodo\footnote{\url{https://zenodo.org/records/15646861}}.

\section{Evaluation}
\label{sec:results}

In this section, we present the particularities of the evaluation framework as well as the metric used and the results. Finally, we also disseminate the computational cost of the benchmarked systems.

\subsection{Evaluation framework}
\label{subsec:evaluation}
\emph{PeakNetFP} evaluation is strictly based on \emph{NeuralFP} to allow a fair comparison and can be examined in the repository accompanying this publication. We train both \emph{PeakNetFP} and \emph{NeuralFP} for $100$ epochs using a batch size of $240$ with Adam optimizer \cite{adam2014} following the authors' recommendation \cite{chang2021neural}.
Table \ref{tab:network_layers} summarizes the parameters of \emph{PeakNetFP} layers, including the number of anchor peaks $N$, the queryballs radii $R$, and the number of peaks per queryball grouping $G$.

\begin{table}[b]
    \centering
    \begin{adjustbox}{max width=0.85\columnwidth}
    \begin{tabular}{@{}lccccccc@{}}
        \toprule
                      &     &     &     &     & \multicolumn{3}{c}{MLP} \\
        \cmidrule{6-8}
        Layer         & $N$ & $j$ & $G$ & $R$ & A & B & C \\ 
        \midrule
        \midrule
        \multirow{3}{*}{SA + MSG 1} & \multirow{3}{*}{200} & 1 & 4  & 0.1 & 16 & 16 & 32  \\
                                    &                      & 2 & 8  & 0.2 & 32 & 32 & 64  \\
                                    &                      & 3 & 16 & 0.3 & 32 & 48 & 64  \\
        \cmidrule{1-8}
        \multirow{3}{*}{SA + MSG 2} & \multirow{3}{*}{100} & 1 & 4  & 0.2 & 32 & 32 & 64  \\
                                    &                      & 2 & 8  & 0.3 & 64 & 64 & 128 \\
                                    &                      & 3 & 16 & 0.4 & 64 & 64 & 128 \\
        \cmidrule{1-8}
        SA                          &                      &   &    &     & 128 & 256 & 128 \\
        \bottomrule
    \end{tabular}
    \end{adjustbox}
    \caption{\emph{PeakNetFP}\ layers and their parameters: number of anchors $N$, layer index $j$, number of peaks to group $G$, grouping radius $R$, and dimensions of the 3 MLP layers $A$, $B$ and $C$.}
    \label{tab:network_layers}
\end{table}

Since there is no public \emph{QuadFP} implementation, we create our own version based on the original paper \cite{sonnleitner2016robust}. Although not all implementation details are provided for exact replication, we closely followed the model's key aspects, such as computing more quads for queries than references and applying cascading heuristics to efficiently filter out irrelevant quads during comparison. We validate our implementation in section \ref{subsec:results} by comparing our results with the ones from the original publication \cite{sonnleitner2016robust}.

We evaluate \emph{PeakNetFP} as well as \emph{QuadFP} and \emph{NeuralFP} in the presence of time stretching ranging from the extreme values 0.5x to 2x the original speed. Additionally, we test each model with query lengths of 2, 3, 5, 6, and 10 seconds to ensure relevance for query-by-example applications. 1s queries are not considered, since their resulting size with time factors over 1 would make them smaller than \emph{NeuralFP} window size.

To compare the AFP systems fairly, we align with the literature \cite{chang2021neural} by using Top-1 hit rate $\text{HR}@1$ defined as the number of hits at Top-1 divided by the number of queries.
Note that both \emph{NeuralFP} and \emph{PeakNetFP} always returns a match, so in this case, Top-1 hit rate is equivalent to both precision and recall. Future work could include training a classifier on the matching scores to adapt the system to out-of-vocabulary queries.

\subsection{Results}
\label{subsec:results}
\begin{figure}[t]
    \centering
    \includegraphics[width=\columnwidth]{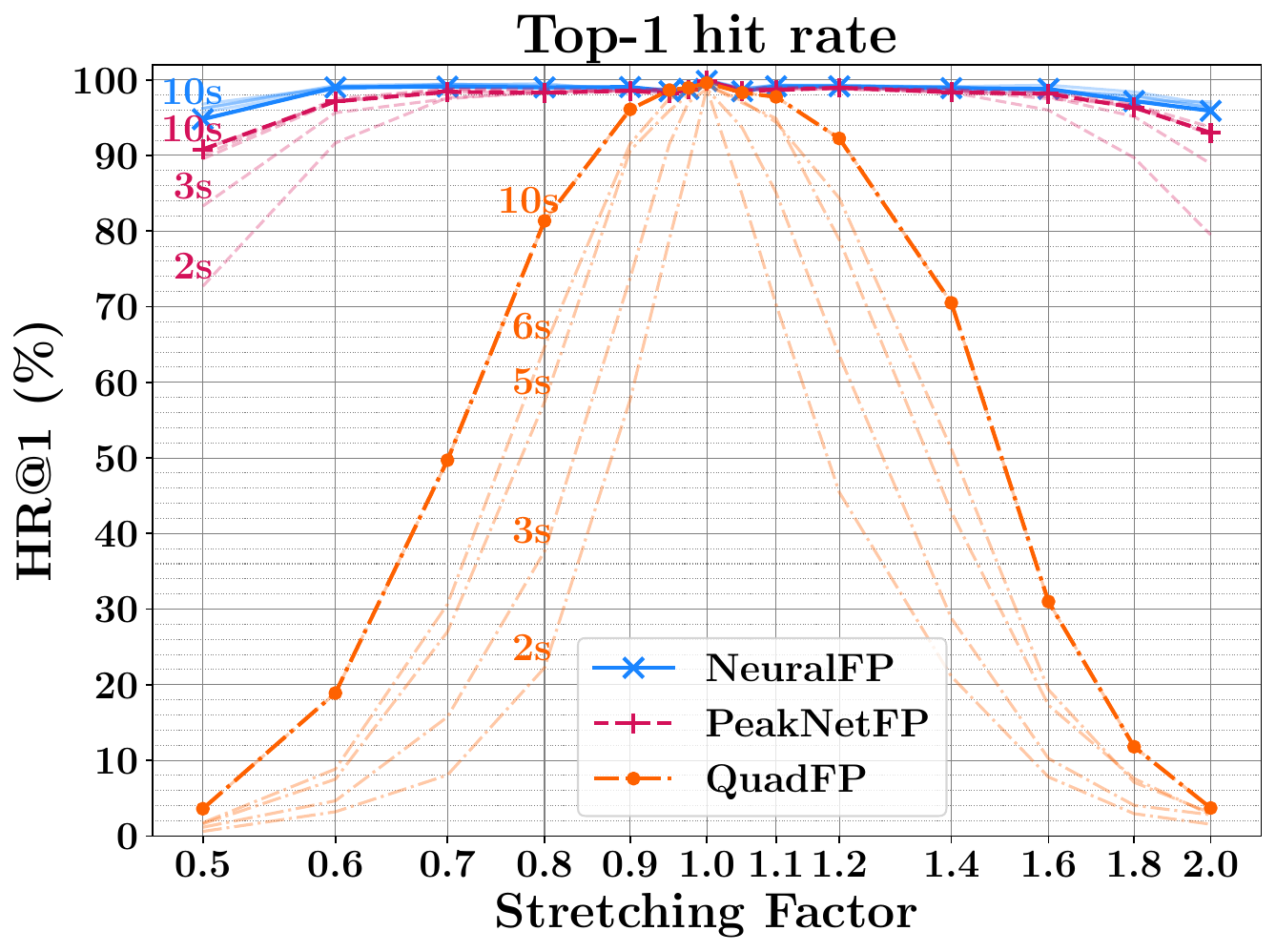}
    \caption{Top-1 hit rate ($\text{HR}@1$) as a function of stretching factors for \emph{NeuralFP}, \emph{PeakNetFP} (ours), and \emph{QuadFP}. Each curve represents one (model, query length) pair.}
    \label{fig:results-top1}
\end{figure}

To check the validity of our custom implementation, we compare our \emph{QuadFP} results with the ones reported on Table I in \cite{sonnleitner2016robust}. Our implementation of \emph{QuadFP} shows better results than the original for short queries ($\leq$5 seconds), with an average 48\% $\text{HR}@1$ for 2s queries in our case against a reported 28\% for 2.5s queries in \cite{sonnleitner2016robust}, and slightly worse results for larger ones, with 94\% for our implementation vs 98\% in \cite{sonnleitner2016robust} for 20-second queries. We acknowledge that the datasets differ between both studies (FMA versus Jamendo), but hypothesize that since their size and nature is similar, the comparison in the context of AFP systems still remains valid. In summary, we conclude that our implementation yields results comparable to \cite{sonnleitner2016robust}, with remaining differences coming from either evaluation datasets or implementation differences.

Figure \ref{fig:results-top1} illustrates Top-1 hit rates $\text{HR}@1$ of all 3 models as a function of stretching factor. For each model, we report 5 different curves that correspond to the 5 query lengths tested, drawn in blue plain lines for \emph{NeuralFP}, orange dashed-dot lines for \emph{QuadFP}, and magenta dashed lines for our model \emph{PeakNetFP}.
We highlight the curves corresponding to 10 seconds queries to compare with the best setup for our \emph{QuadFP} baseline.

\emph{PeakNetFP} outperforms \emph{QuadFP} globally, effectively handling time stretching. It also exhibits excellent performance, achieving over 98\% HR@1 for 10-second queries within the commonly reported 0.7 to 1.4 stretching factor range\cite{sonnleitner2016robust,yao2018enhancing,son2020robust}. For extreme stretching factors (<0.7 and >1.4) \emph{PeakNetFP} performance decreases a bit but still maintains over 90\% $\text{HR}@1$. As a reference, \emph{QuadFP} only achieves 3.6\% $\text{HR}@1$ for a 0.5 factor.
In fact, we observe that \emph{QuadFP} performs well for minor stretching (0.9 to 1.1) as reported before \cite{sonnleitner2016robust}, but its performance rapidly deteriorates as the stretching deviates further from 1, reaching nearly zero $\text{HR}@1$ at extreme stretching factors of 0.5 and 2. This effect is probably due to the lack of enough preserved quads at such strong time stretching factors. In terms of query length, \emph{QuadFP}'s performance degrades quickly as queries get smaller. It should be reminded here that \emph{QuadFP} is a rule based algorithm that does not require training as opposed to \emph{PeakNetFP} or \emph{NeuralFP}.

Results on the SOTA model \emph{NeuralFP} exhibit a strong robustness to time stretching, even in extreme cases. As a reminder, \emph{NeuralFP} processes entire spectrograms while \emph{PeakNetFP} processes sparse peaks only. Nonetheless, for the commonly reported time stretching factors (0.7 to 1.4), \emph{PeakNetFP} and \emph{NeuralFP} obtain almost identical performance, with a maximum difference of $\pm0.7$\% $\text{HR}@1$.
For the more extreme factors, both systems performance is diminished, with \emph{PeakNetFP} being more affected than \emph{NeuralFP}, with the maximum difference between systems being 1.85\% $\text{HR}@1$ at factor 0.5. Regarding the query length, \emph{PeakNetFP} requires 5s queries at least to keep performance systematically above 0.9 at extreme time factors.

We conclude that \emph{PeakNetFP} has a performance comparable to \emph{NeuralFP} with a slight decrease at extreme stretching factors. 
However, \emph{PeakNetFP} is significantly lighter than \emph{NeuralFP}. It uses an input of 256 3D peaks, approximately 11 times smaller than \emph{NeuralFP}'s $256 \times 32$ spectrograms. With 169k trainable parameters, \emph{PeakNetFP}'s model size is 100 times smaller than \emph{NeuralFP}'s 16.9M. This also requires substantially less inference memory: 800MiB for \emph{PeakNetFP} versus 2338 MiB for \emph{NeuralFP} (batch size 125 on a single RTX 3090). This improves our model's scalability and reduces memory usage for catalog embedding generation.

\section{Conclusion}
\label{sec:conclusion}

In this work, we introduce a novel audio fingerprinting system, \emph{PeakNetFP}, that is designed as a hybrid approach combining the strengths of traditional peak-based fingerprint systems, heavily used in industrial contexts, with modern neural network-based representation learning approaches. We use a computer vision-inspired point cloud network to handle sparse peaks, which we use in a contrastive learning approach similar to modern AFP methods. Our evaluation in the context of extreme time stretching demonstrates that \emph{PeakNetFP} consistently outperforms the SOTA on time-stretched data, \emph{QuadFP}. Moreover, we show that, while the spectrogram-based SOTA in AFP \emph{NeuralFP} performs very well in such a task, our model \emph{PeakNetFP} can achieve comparable performance while working on peaks and thus requiring 11 times smaller input data, and using 100 times less parameters than the former.

In conclusion, \emph{PeakNetFP} provides a scalable and efficient solution for audio identification tasks that involve significant tempo alterations, combining the compactness of peak-based methods with the robustness and flexibility of neural networks. It improves with respect to traditional methods for severe to extreme stretching factors, and appears as an alternative to fully neural approaches, especially for contexts where memory and computational efficiency are critical. Future works will focus on improving the model for applications beyond time stretching, such as pitch shifting.

\section{Acknowledgements}
This research is part of \emph{resCUE – Smart system for automatic usage reporting of musical works in audiovisual productions (SAV-20221147)} funded by CDTI and the European Union - Next Generation EU, and supported by the Spanish Ministerio de Ciencia, Innovaci\'on y Universidades and the Ministerio para la Transformaci\'on Digital y de la Función P\'ublica. Furthermore, it has received support from the Industrial Doctorates plan of the Secretaria d’Universitats i Recerca, Departament d’Empresa i Coneixement de la Generalitat de Catalunya, grant agreement No. DI46-2020.

\bibliography{main}

\end{document}